\documentclass[11pt,letterpaper]{article}

\usepackage{jheppub}
\usepackage{color}
\usepackage{graphicx}
\usepackage{wrapfig}

\usepackage{verbatim}
\usepackage{amsmath}
\usepackage{amssymb}
\usepackage{subfig}
\usepackage{url}
\usepackage{bbold}
\usepackage{xspace}

\usepackage{multirow}
\usepackage{threeparttable}
\usepackage{paralist}

\usepackage{color}
\definecolor{darkgreen}{rgb}{0,0.5,0}

\newcommand{\GeV}{\text{GeV}}
\newcommand{\TeV}{\text{TeV}}

\newcommand{\U}{\text{U}}
\newcommand{\SU}{\text{SU}}

\DeclareRobustCommand{\Sec}[1]{Sec.~\ref{#1}}

\DeclareRobustCommand{\Fig}[1]{Fig.~\ref{#1}}
\DeclareRobustCommand{\Figs}[2]{Figs.~\ref{#1} and \ref{#2}}
\DeclareRobustCommand{\Eq}[1]{Eq.~(\ref{#1})}

\newcommand{\be}{\begin{equation}}
\newcommand{\ee}{\end{equation}}

\newcommand{\mb}[1]{\boldsymbol{#1}}

\begin{document}

\title{Modified Higgs Sectors and NLO Associated Production}
 
\author[a]{Christoph Englert}
\author[b]{and Matthew McCullough}

\affiliation[a]{Institute for Particle Physics Phenomenology,
  Department of Physics,\\Durham University, Durham DH1 3LE, UK}
\affiliation[b]{Center for Theoretical Physics, Massachusetts
  Institute of Technology,\\Cambridge, MA 02139, USA}

\emailAdd{christoph.englert@durham.ac.uk}
\emailAdd{mccull@mit.edu}

\date{\today}

\abstract{}


\arxivnumber{}

\preprint{DCPT/13/10,~IPPP/13/20,~MIT-CTP {4441}}

\abstract{Many beyond the Standard Model (BSM) scenarios involve Higgs
  couplings to additional electroweak fields.  It is well established
  that these new fields may modify $h \to \gamma\gamma$ and $h \to
  \gamma Z$ decays at one-loop.  However, one unexplored aspect of such
  scenarios is that by electroweak symmetry one should also expect
  modifications to the $h Z Z$ coupling at one-loop and, more
  generally, modifications to Higgs production and decay channels
  beyond tree-level.  In this paper we investigate the full BSM
  modified electroweak corrections to associated Higgs production at
  both the LHC and a future lepton collider in two simple SM
  extensions. From both inclusive and differential NLO associated
  production cross sections we find BSM-NLO corrections can be as
  large as ${\cal{O}}(\gtrsim 10\%)$ when compared to the SM
  expectation, consistent with other precision electroweak
  measurements, even in scenarios where modifications to the Higgs
  diphoton rate are not significant.  At the LHC such corrections are
  comparable to the involved QCD uncertainties.  At a lepton collider
  the Higgs associated production cross section can be measured to
  high accuracy (${\cal{O}} (1 \%)$ independent of uncertainties in
  total width and other couplings), and such a deviation could be
  easily observed even if the new states remain beyond kinematic
  reach.  This should be compared to the expected accuracy for a
  model-independent determination of the $h\gamma\gamma$ coupling at a
  lepton collider, which is ${\cal{O}} (15 \%)$.  This work
  demonstrates that precision measurements of the Higgs associated
  production cross section constitute a powerful probe of modified
  Higgs sectors and will be valuable for indirectly exploring BSM
  scenarios.}

\maketitle

\section{Introduction}
\label{sec:introduction}
Last year the ATLAS and CMS collaborations discovered a new scalar in
the bosonic Higgs search channels \cite{atlas:2012gk,cms:2012gu}
exhibiting properties consistent with the Standard Model (SM) Higgs
boson with mass $m_h\simeq 125~\GeV$.  This mass is fortuitous from
both experimental and theoretical perspectives since multiple Higgs
decay channels can potentially be observed, allowing for a
multi-facetted exploration of electroweak symmetry breaking.  The rich
experimental possibilities, combined with the fact that the Higgs is
the focus of many new physics scenarios, propels the Higgs to center
stage in the effort to connect BSM theory with experiment.  This leads
to a number of phenomenologically important questions. If there is new
physics, what is its nature?\footnote{Due to early tentative hints for an excess in $h\to \gamma\gamma$ rates
 this question has stimulated a great deal of theoretical activity.  See e.g.\ \cite{hgamgam,Joglekar:2012vc} for explicit models and
  also discussions of the implications of BSM contributions to the
  $h\gamma\gamma$ coupling.  These hints have persisted in recent ATLAS analyses \cite{ATLAS_moriond}, but only at the level of $2 \sigma$, and CMS have not yet released an updated analysis \cite{CMS_moriond}, so the status of this excess is very uncertain and we do not consider it a motivating factor in this work.
}  How might we hope to observe it directly
or indirectly?  Addressing the latter question requires understanding
in detail how new physics scenarios might affect Higgs production and
decay properties and, consequently, precision analyses at present and
future colliders.  Even if no significant deviations in the Higgs
properties are observed from the first run of the LHC, these issues
remain at the top of the phenomenological agenda.

\begin{figure}[!t]
  \centering
  \subfloat[]{\label{fig:hgamgam}
    \includegraphics[height=1.4in]{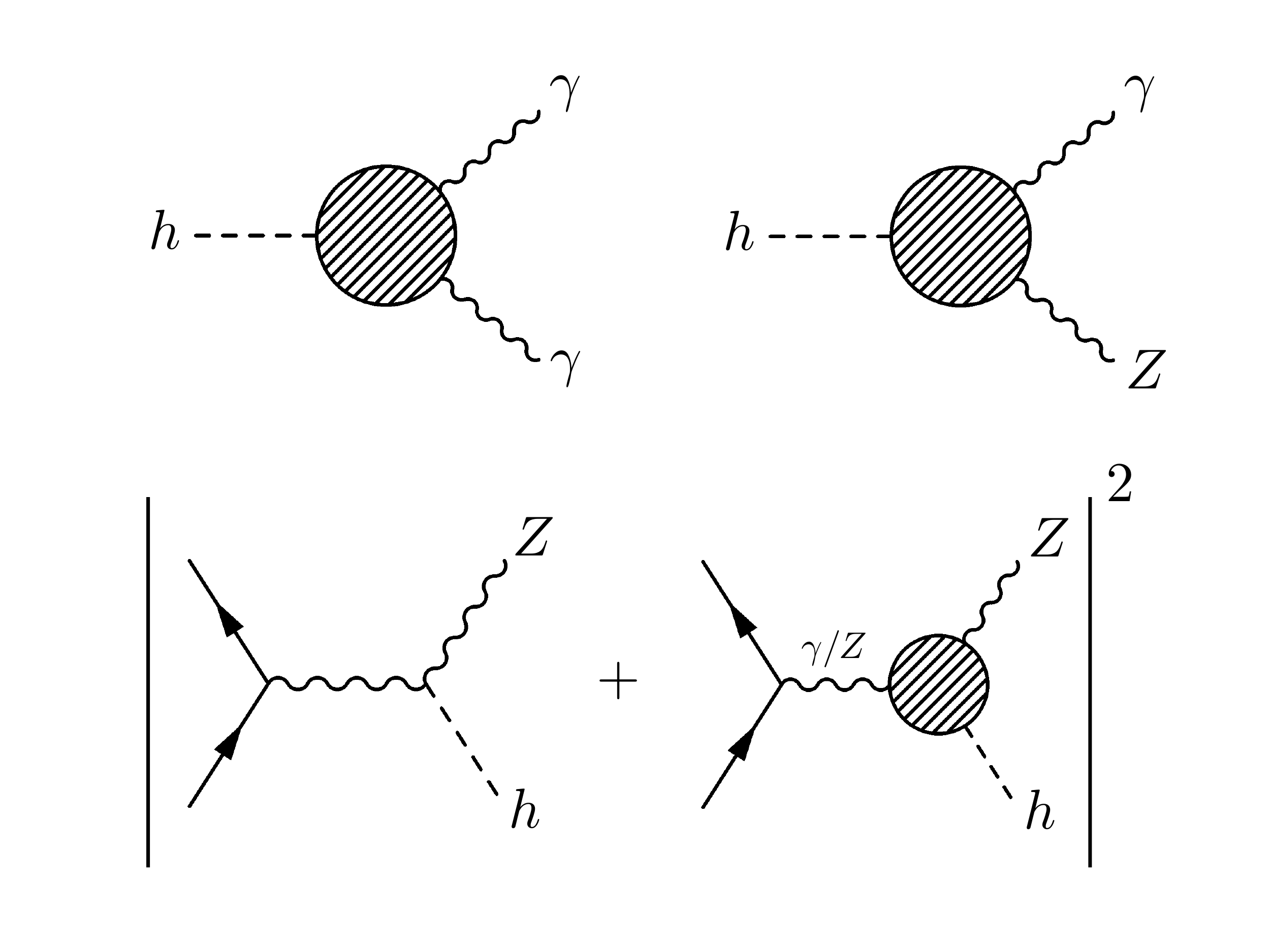}}   
  \hspace{0.6in}
  \subfloat[]{\label{fig:hgamZ}
    \includegraphics[height=1.4in]{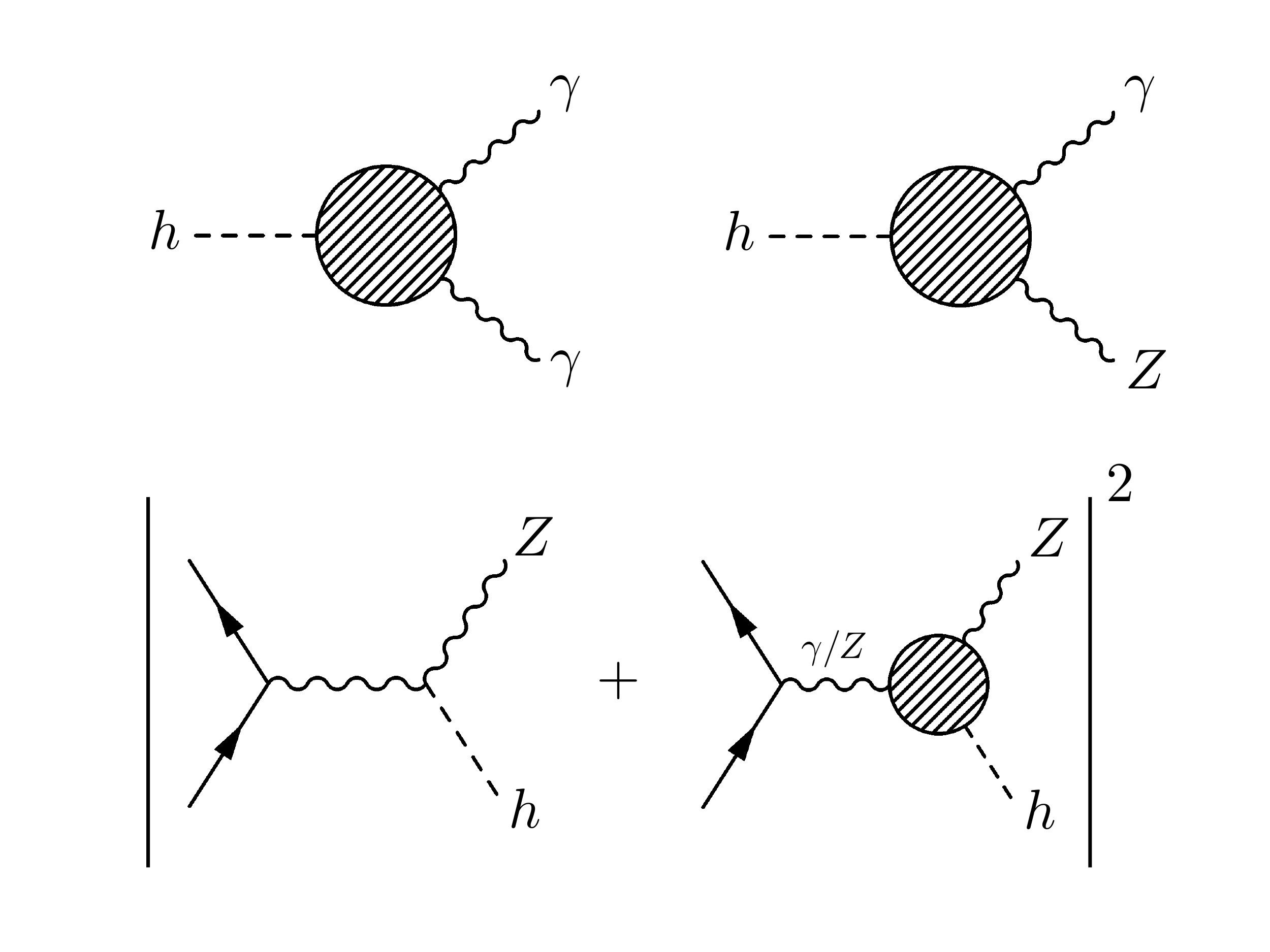}}
  \caption{New electroweak-charged fields coupled to the Higgs
    contribute at one loop to (a) $h\rightarrow \gamma \gamma$ decays
    and (b) $h \rightarrow \gamma Z$ decays.  The branching ratios to
    these final states are sensitive to the total Higgs width, which
    depends on Higgs couplings to other SM and BSM fields.  The total
    rate in these channels also depends on the Higgs production cross
    section.  Thus experimental determination of the $h\gamma\gamma$ and
    $h \gamma Z$ couplings, either at the LHC or a lepton collider, is
    subject to uncertainties in all of the Higgs couplings.  For these
    reasons, at a $250$~GeV lepton collider with $250 \text{ fb}^{-1}$
    integrated luminosity the $h \gamma \gamma$ vertex can be determined
    to an accuracy of $\mathcal{O} (15\%)$ \cite{Klute:2013cx}.}
  \label{fig:hgamgamhgamZ}
\end{figure}

\begin{figure}[!t]
  \centering
  \includegraphics[height=1.4in]{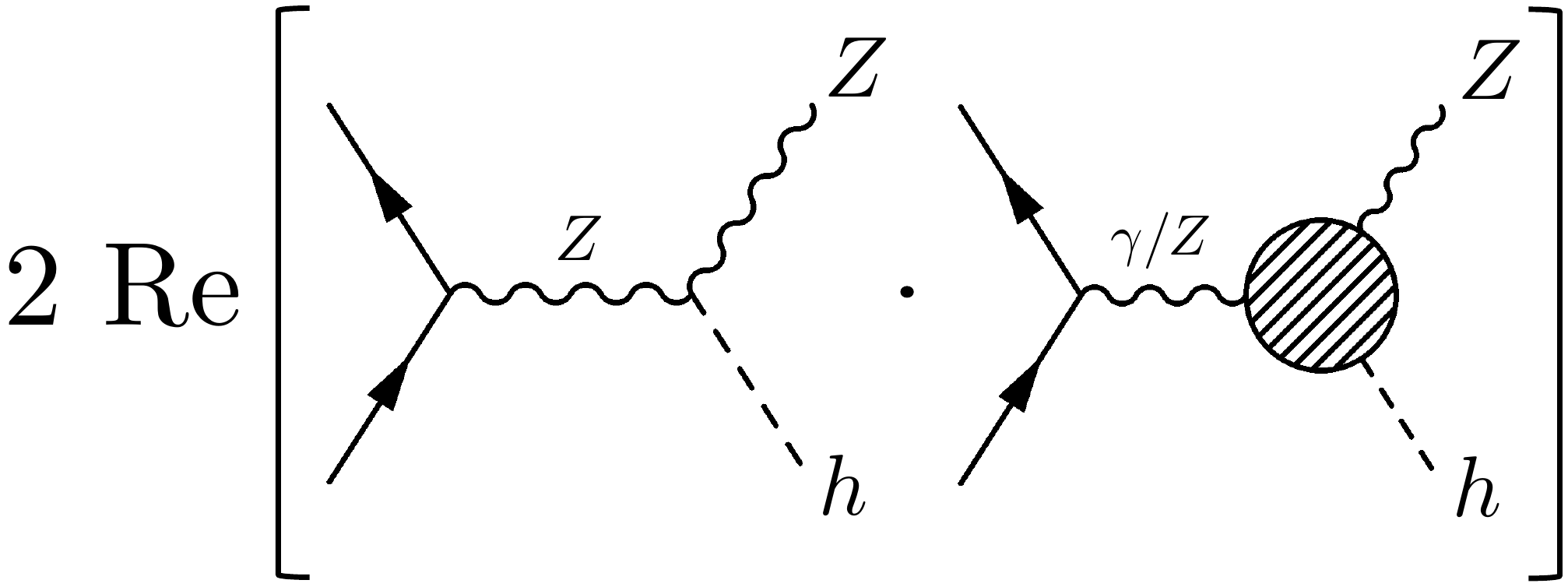} 
  \caption{Higgs associated production.  New electroweak-charged
    fields coupled to the Higgs typically contribute to this amplitude
    at one loop, interfering with the tree-level amplitude.  The $hZZ$
    coupling can be measured to a high degree of accuracy
    ($\mathcal{O} (1\%)$ \cite{Klute:2013cx}) at a lepton collider by
    measuring $Z$-recoils in the inclusive associated production
    process, regardless of uncertainties regarding the total Higgs
    width.  Hence a lepton collider gains better sensitivity to the
    $hZZ$ coupling than to the $h \gamma \gamma$ and $h \gamma Z$
    couplings, and opens up the possibility to discover new
    physics through anomalous contributions to the
    $hZZ$ vertex.}
\label{fig:assocprod}
\end{figure}

Perturbative BSM scenarios which involve Higgs couplings to new
electroweak fields are often characterized by new states charged under
$\SU(2)_L\times \U(1)_Y$ which obtain, at least partially, their mass
from the Higgs sector.  In
Figs.~\ref{fig:hgamgam},~\ref{fig:hgamZ},~and~\ref{fig:assocprod}, we
schematically demonstrate that by electroweak symmetry Higgs
$h\to\gamma \gamma$ decays, $h\to\gamma Z$ decays, and associated
production at the Tevatron, LHC, and a future lepton
collider\footnote{We will use the term `lepton collider' in reference
  to any future high energy $e^+e^-$ synchrotron, linear collider or
  Higgs factory.} are typically influenced by such couplings. In
particular, new physics contributions that enter $h\to \gamma\gamma$
and $h \to Z\gamma$ at leading order are directly correlated with the
next-to-leading order (NLO) electroweak corrections to Higgs boson
production in associated production.\footnote{New electroweak fields
  will also contribute in the $Z$ self-energy and $\gamma/Z$ mixing.
  We include these contributions in all calculations, but not in
  diagrams.  Similar corrections will also enter into $h \rightarrow Z
  Z^\star$ decays.  We do not consider this here, although it is
  certainly deserving of further investigation.} This observation is
not merely a technical curiosity, but is relevant across the broad
scope of the experimental Higgs frontier since the associated
production process lies at the heart of a number of critical
measurements, which we now briefly review.\footnote{For other
  BSM applications of the associated Higgs production process
  see e.g.\ \cite{Kilian:1996wu,Ellis:2013ywa}.}
  
While there are emerging hints of $h\to \tau\tau$ at the LHC
\cite{atlas:conf,cms:conf} and $h\to b\bar b$ at the
Tevatron~\cite{TEVNPH:2012ab}, the Higgs boson candidate remains to be
discovered in the fermionic final states. Given that the decay $h\to
b\bar b$ in the SM for $m_h\simeq 125~\GeV$ is the largest decay
channel, with a branching ratio $\text{BR}_{SM}(h\to b\bar b)\simeq
60\%$, not observing the Higgs boson in this channel indirectly
induces a large uncertainty on the extraction of Higgs couplings to
other fields.  We note however that this uncertainty can be
artificially decreased by making assumptions on the total width of the
particle.  Looking to the future, observing $h\to b\bar b$ at the LHC
in either gluon fusion or weak boson fusion will be difficult, due to
overwhelmingly large QCD backgrounds, trigger issues, or the involved
selection criteria which in the latter search channel typically
involve central jet vetoes
\cite{Dokshitzer:1991he,Barger:1994zq,atlas:2012gk,cms:2012gu}.\footnote{Although
  difficult, we note that observation of $h\to b\bar b$ in the weak
  boson fusion channel is possible in the future.  We thank Markus
  Klute for bringing this to our attention.} However, given the
success of jet substructure techniques on boosted final states from
associated Higgs production $pp\to hZ$
\cite{Butterworth:2008iy,Soper:2010xk}, there is good reason to
believe that such an analysis will shed light on $h\to b\bar b$ in the
future \cite{Chatrchyan:2012ww,:2012zf,ajr}.  Thus good theoretical
control over the associated production cross section is integral to
future determination of the fermionic Higgs couplings at the
LHC.\footnote{It should also be noted that one may be able to study
  different production cross sections with the same Higgs decays.  For
  example, $h\to WW^\ast$ decays may be observable separately in gluon
  fusion, vector boson fusion, and associated production channels (see
  {\it{e.g.}}\ \cite{CMS-PAS-HIG-12-039,CMS-PAS-HIG-12-042}) leading
  to relations between cross sections.}

The status of the associated production process is elevated further at
any future lepton collider since at low energies ($\sim 250$~GeV)
associated production is the dominant Higgs production process and
influences all Higgs observables.  Furthermore, measurements of the
inclusive associated production cross section will allow determination
of the $hZZ$ coupling to ${\cal{O}} (1\%)$ accuracy, regardless of the
total Higgs width, making this the most accurately determined of all
the Higgs couplings.  The associated production cross section also has
ramifications for precision determination of a number of Higgs
couplings as the total Higgs decay width can be inferred through a
combination of the inclusive associated production cross section,
specifically $Z\to\mu^+ \mu^-$ recoiling against the Higgs, and other
production and decay channels.  Hence, if there are any consequences
of a modified Higgs sector for associated production, they will be
important at lepton colliders.
  
Non-decoupling parts of the one-loop contributions of
Figs.~\ref{fig:hgamgam}~and~\ref{fig:hgamZ} are sensitive to physics
beyond the SM even if the new mass scale lies well above the LHC or
lepton collider energy reach. If the new physics mass scale $\Lambda$
is large enough, it is customary to parametrize deviations from the SM
by including higher dimensional operators that arise from integrating
out the dynamics above $\Lambda$ in an effective field theory
language.\footnote{For early work in the gauge and Higgs sectors see
  Refs.~\cite{Hagiwara:1986vm,Buchmuller:1985jz}} This approach has
the benefit of being model independent, but is limited due to the very
nature of effective field theory. This does not only follow from the
non-renormalizable character of effective field theory, which
systematically hinders theoretical precision \cite{Passarino:2012cb},
but sheer experimental reality puts the link between experiment and
theoretical interpretation under stress.  Many well-motivated physics
extensions, especially in the electroweak sector, result in a
phenomenology that can be consistent with current data without having
a too high new physics scale $\Lambda\sim 100~\GeV$. In this case, a
serious drawback of the effective theory approach is that experimental
signal versus background discriminating selection criteria typically
need to be designed to probe momentum transfers $p_T\sim \Lambda$
simply in order to overcome the huge SM backgrounds. This means we
would probe the effective theory at scales at which it cannot be
considered valid. An example for such an analysis is again the boosted
$h\to b\bar b$ associated production where $p_T(Z)\gtrsim 120~\GeV$
facilitates both triggering the event and rejecting the large $t\bar
t$+jets background \cite{Butterworth:2008iy,Soper:2010xk}. Therefore
the precision analysis of SM extensions at similar energy scales can
become model-dependent and only dedicated calculations allow one to
draw a qualitatively complete picture.\footnote{Note also that by
  typically probing a large center-of-mass energy the information
  encoded in electroweak precision observables such as the $S,T,U$
  observables \cite{Peskin:1990zt,Peskin:1991sw} do not govern the
  collider phenomenology.}

For the aforementioned reasons we choose to study particular models,
since although one loses some degree of generality, this enables
precision calculations and also serves to illustrate the features one
might expect in a broader range of SM extensions.  To determine where
the new electroweak physics may become apparent, in this paper we
investigate the correlation between two obvious observables in which
new electroweak physics may lead to modifications at one-loop:
corrections to $h\to \gamma \gamma$ decays, which is very commonly
studied, and NLO electroweak corrections to associated Higgs
production at the LHC and a future lepton collider where this
cross section can be measured to high precision.  To do this we employ
two concrete models. We first discuss a vector-like leptonic fourth
generation in Sec.~\ref{sec:VL4} and then a charged scalar Higgs
sector extension of the SM in Sec.~\ref{sec:Scalars}. For both models
we establish consistency with current constraints in terms of the
precision electroweak $S,T,$ and $U$, parameters
\cite{Peskin:1990zt,Peskin:1991sw}.

Simple electroweak extensions
typically involve singlets under Quantum Chromodynamics (QCD)
$\SU(3)_C$. Depending on its spin, color-charged matter can compete
with SM electroweak fields at one-loop, but can also be directly
constrained at the LHC due to limits on the production of colored
particles~\cite{ATLAS-CONF-2012-148}.  Discarding non-trivial color
assignments, which we will do in the following, does result in some
degree of qualitative theoretical deficiency, however if one wished to
include colored matter then, by considering \Fig{fig:hgamgam},
\Fig{fig:hgamZ}, and \Fig{fig:assocprod}, it is clear that the BSM
amplitude simply scales uniformly with the casimir of the $\SU(3)_C$
matter representation.

In Sec.~\ref{sec:lincol} we consider electroweak corrections at lepton
colliders, at which the impact of the electroweak quantum effects is
most pronounced and their observation is challenged least.  We set the
stage in the context of the SM in Sec.~\ref{sec:SM}, bridging our work
to the existing literature. In Sec.~\ref{sec:BSM} we discuss
modifications of the NLO electroweak phenomenology in the presence of
BSM interactions. For completeness we provide a survey of the expected
BSM NLO electroweak effects at the LHC in Sec.~\ref{sec:radLHC},
utilizing exemplary parameter points that are motivated from our
lepton-collider discussion of Sec.~\ref{sec:BSM}.  We summarize this
work and give our conclusions in Sec.~\ref{sec:Conclusions}.

\section{Example BSM Scenarios}
\label{sec:sample}

\subsection{Vector-Like Fourth Generation Leptons}
\label{sec:VL4}
As a first illustrative BSM scenario we discuss a model with an
additional family of vector-like leptons.\footnote{In
  Ref.~\cite{Joglekar:2012vc} the model has been discussed in the
  context of enhanced $h\to \gamma\gamma$ decays, dark matter and
  constraints from direct searches.} The model introduces a full
vector-like generation of leptons and the Lagrangian, suppressing the
kinetic and gauge interactions, is\footnote{PMNS-like mixing of the
  leptons with the first three generations is quite constrained and
  can be avoided by imposing global symmetries. This is implicitly
  assumed in the following. We borrow the conventions of
  Ref.~\cite{Joglekar:2012vc}.}
\begin{subequations}
  \label{eq:4llag}
  \begin{align}
    \label{eq:vector}
    -{\cal L}\,\, &\supset m_\ell \bar{\ell}'_L {\ell }''_R + m_e
    \bar{e}''_L {e}'_R
    + m_\nu \bar{\nu}_L'' {\nu}_R' + \rm{h.c.}   \\
    \label{eq:higgs}
    & +Y_c' (\bar{\ell}'_L H ) {e}_R' + Y_n' (\bar{\ell}'_L i\sigma^2
    H^\dagger) {\nu}_R' + Y_c'' (\bar{\ell}''_R H ) {e}_L''
    + Y_n'' (\bar{\ell}''_R i\sigma^2 H^\dagger)  {\nu}_L'' + \rm{h.c.}
  \end{align}
\end{subequations}
where $\ell'_L, \ell''_R=(\mb{2},-1/2)$, $e''_L,e'_R=(\mb{1},-1)$, and
$\nu''_L,\nu'_R=(\mb{1},0)$ under $\SU(2)_L\times \U(1)_Y$.  The
left-right symmetric character of this extension guarantees
cancellation of anomalies.

The parameter space of this model is large, and so we choose to make
some restrictions to facilitate exploration of the relevant
phenomenology.  In the following we assume all vector-like
masses take a common value 
\begin{subequations}
  \begin{equation} 
    \label{eq:ident}
    m_\ell=m_e=m_\nu = m_V ~~, 
  \end{equation}
  where the subscript $V$ simply denotes that this mass contribution
  is vector-like.  We also assume that all charged fields have the
  same Yukawa couplings and obtain the same mass contributions from
  the Higgs.  We make a similar simplification for the neutral fields
  but allow their Yukawa couplings to differ from the charged fields.
  Thus we define
  \begin{align}
  Y_c' v_h/ \sqrt{2} & = Y_c'' v_h/ \sqrt{2} = m_{Ch} \\
  Y_n' v_h/ \sqrt{2} & = Y_n'' v_h/ \sqrt{2} = m_{Ch}+\Delta_\nu ~~,
  \label{eq:VL4hcoup}
\end{align}
\end{subequations}
where the subscript $Ch$ denotes that this mass contribution is chiral
in nature.  This simplified version of a vector-like fourth generation
of leptons is hence fully specified by the parameter set
$\{m_{Ch},m_V,\Delta_\nu \}$.

Finally we define $m_{E_1}$ to be the mass of the lightest charged
fermion and throughout will assume $m_{E_1} > 125$~GeV, just beyond
the direct production reach of a $250$~GeV lepton collider.

The mass eigenstate basis that results from diagonalizing
\eqref{eq:4llag} with (bi)unitary transformations generates mixings in
the gauge and Higgs interactions \eqref{eq:higgs}.  While these
mixings are important for processes involving production and decay of
the new fields, they will also typically leave footprints in the low
energy physics, below the energy scales at which the new physics is
directly accessible.  While the focus in this work is on modifications
of Higgs physics at scales $\Lambda \sim m_h$, we must also consider
the effects on physics at scales $\Lambda \sim m_Z$.  Such
modifications are conventionally expressed in terms of the precision
electroweak $S,T$, and $U$, parameters
\cite{Peskin:1990zt,Peskin:1991sw}, which are sensitive to the degree
of the model's non-chirality and any additional custodial isospin
violation. For our parameter identifications
Eqs.~\eqref{eq:ident}-\eqref{eq:VL4hcoup}, the former is controlled by
$m_{Ch}$ and the latter by $\Delta_\nu$.  In the language of effective
field theory, $U$ follows from a dimension eight operator as opposed
to the dimension six operators which parametrize $S$ and $T$, and
hence is typically suppressed \cite{Barbieri:2004qk}.  In this model
contributions to $S$ and $T$ can be considerable, and so when
considering the influence of the new fields on Higgs associated
production and decays we will also include precision electroweak
constraints throughout.\footnote{Due to the sterile neutrino
  extension, the model also allows additional Majorana mass terms,
  which can lead to a good dark matter candidate. For the present
  context, {\it i.e.} the ILC and LHC collider phenomenology at NLO,
  the only impact of the Majorana masses is a modified mixing of the
  gauge interactions, which has no effect on the observed $h\to
  \gamma\gamma$, and we do not consider Majorana mass terms in the
  following.}

\subsection{New Electroweak Scalars}
\label{sec:Scalars}
As a second illustrative BSM scenario we consider a model containing
additional scalars.  In particular we include an extra scalar doublet,
$\phi$, with the same quantum numbers as the Higgs doublet.  For the
sake of simplicity we assume $\phi$ is charged under an approximate
global $\U(1)$ symmetry\footnote{We assume this global symmetry is
  broken by some small amount to allow for $\phi$ to decay, with the
  proviso that the operator responsible for this is small enough to be
  irrelevant for the loop corrections we discuss.} which restricts the
scalar potential to contain the following terms
\be 
V \supset m^2_\phi
|\phi|^2 + \lambda |H|^2 |\phi|^2 + \lambda' |H\cdot \phi^\dagger|^2
~~.
\label{eq:scpot}
\ee 
After electroweak symmetry breaking the Higgs vev leads to a mass
splitting between the neutral and charged scalars, $\phi_0$ and
$\phi_+$ and also generates trilinear interactions between the Higgs
and pairs of scalars.  It is useful to trade the parameters of the
above Lagrangian, $\{m_\phi, \lambda, \lambda'\}$, for the more
intuitive set $\{m_{\phi_+},A_{\phi_+},\Delta_\phi \}$, where
$m_{\phi_+}$ is the mass of the charged scalar, $A_{\phi_+}$ is the
trilinear coupling between the Higgs and charged scalar $\mathcal{L}
\supset A_{\phi_+} h |\phi_+|^2$, and $\Delta_\phi$ is the mass
splitting between the neutral and charged scalars
\be \Delta_\phi = m_{\phi_0} - m_{\phi_+}
~~.  
\ee 
As for the vector-like lepton model we will also include precision
electroweak constraints on this model throughout and assume
$m_{\phi_+} > 125$~GeV, such that the lightest charged scalar is
beyond direct production reach at a $250$~GeV lepton collider.

\section{(B)SM Electroweak Radiative Corrections at a Lepton Collider}
\label{sec:lincol}
Although the Higgs was only discovered recently, the program to
understand and calculate radiative corrections in various Higgs
production and decay processes was initiated long ago.  Measuring
electroweak radiative corrections typically requires a precision
machine, so in this section we focus on Higgs production at a lepton
collider, and in particular consider an $e^+ e^-$ collider operating
at $\sqrt{s} = 250$~GeV, for which the dominant production mechanism
is associated production, $e^+ e^- \rightarrow hZ$
\cite{Kilian:1995tr,Kilian:1996wu}. Radiative corrections to
associated production at a lepton collider were calculated some time
ago
\cite{Fleischer:1982af,Jegerlehner:1983bf,Fleischer:1987zv,Denner:1992bc},
and before proceeding to BSM scenarios we revisit these calculations
armed with known values for the top quark and Higgs mass.

\begin{figure}[!t]
  \centering
  \subfloat[]{\label{fig:smtotcross}\includegraphics[height=2.8in]{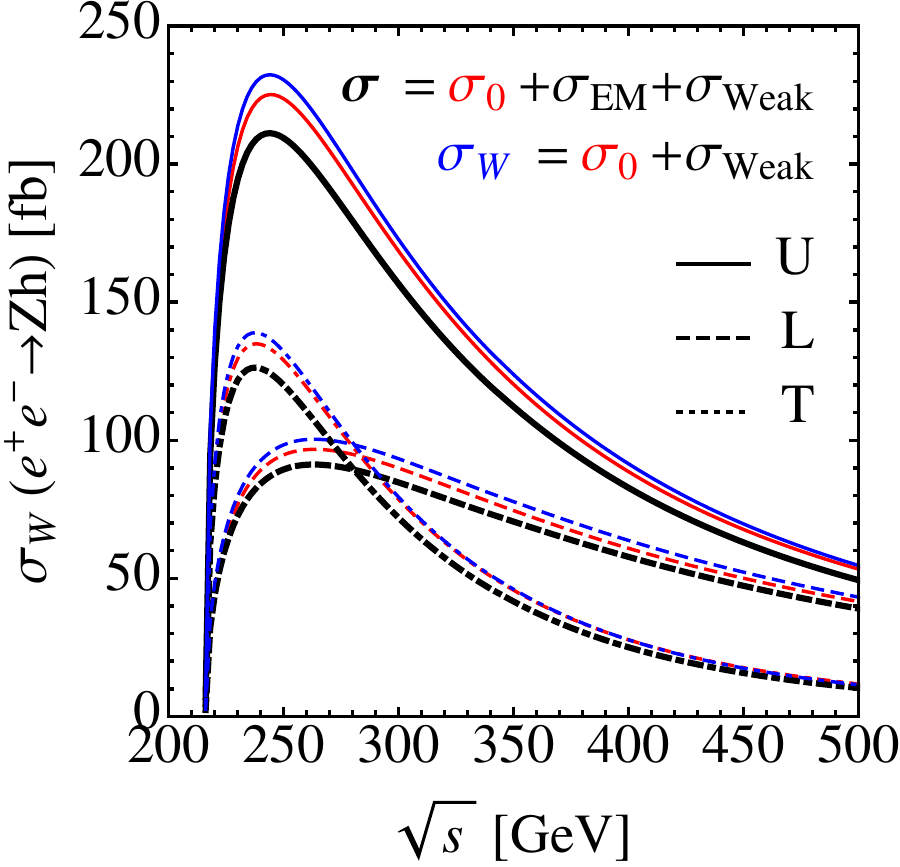}}   
  \hspace{0.1in}
  \subfloat[]{\label{fig:smdiff}\includegraphics[height=2.8in]{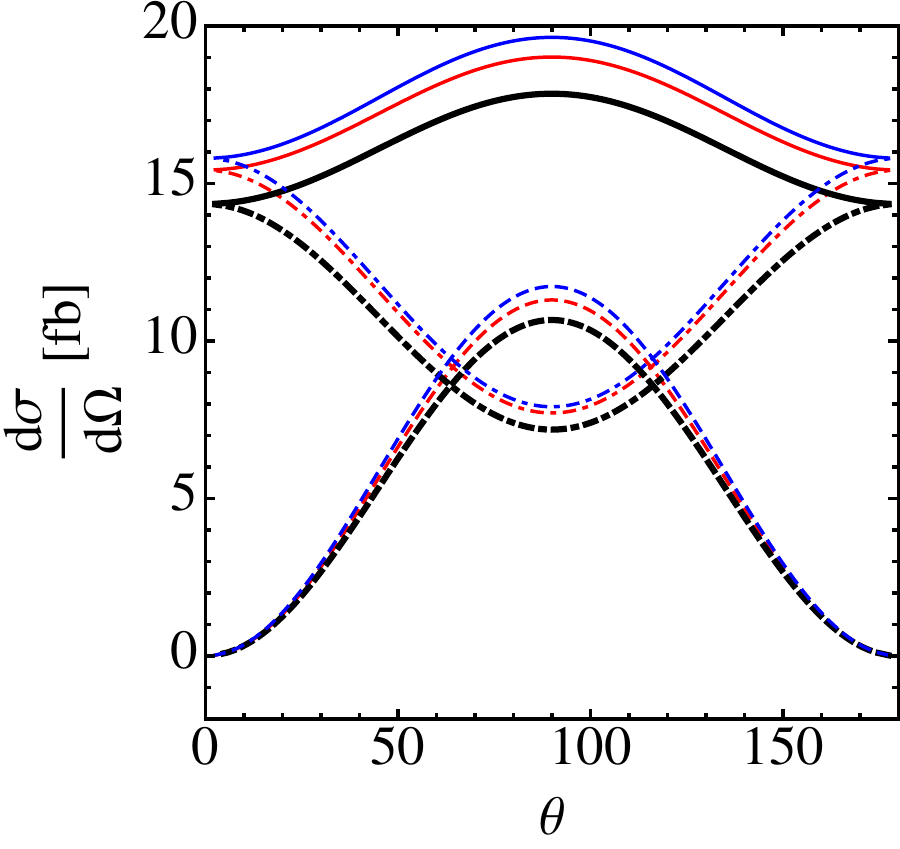}}
  \caption{SM electroweak one-loop corrections to the associated
    production process at a lepton collider.  (a) Total associated
    production cross section at a linear collider as a function of
    $\sqrt{s}$ and (b) differential cross section at $\sqrt{s} =
    250$~GeV.  In both panels the tree-level result is plotted in red
    (middle of a set of three), the result with only weak corrections
    included in blue (top of three), and the full one-loop result in
    black (bottom of three).  The various $Z$-boson polarizations are
    plotted in continuous (unpolarized), dashed (longitudinal) and
    dotted (transverse) lines.}
  \label{fig:SMtotsig}
\end{figure}

\subsection{SM}
\label{sec:SM}

Electroweak radiative corrections to $e^+ e^- \rightarrow hZ$ (and
also $q\bar q\rightarrow hZ$) consist of self energy corrections,
vertex corrections and box diagrams.  We follow \cite{Denner:1992bc}
and calculate these corrections using the {\sc{FeynArts}},
{\sc{FormCalc}}, and {\sc{LoopTools}} packages
\cite{Hahn:2000kx,Hahn:1998yk}.  We calculate counter-terms with
dimensional regularization in the complete on-mass-shell
renormalization scheme
\cite{Denner:1991kt,Aoki:1980ix,Aoki:1980hh,Aoki:1982ed} and also
split the radiative corrections into two classes, electromagnetic and
weak, denoting the full associated production cross section
\be 
\sigma = \sigma_0 + \sigma_{EM}+\sigma_{Weak} ~~,
\ee 
where $\sigma_0$ is tree-level result, $\sigma_{Weak}$ contains the
weak corrections which do not involve photons within loops, and
$\sigma_{EM}$ includes the virtual photon correction to the $eeZ$
vertex as well as soft photon emission.  Including a massless photon
in the virtual contributions of $\sigma_{EM}$ we need to be inclusive
on the final state to guarantee the cancellation of infrared
singularities according to the Kinoshita-Lee-Nauenberg theorem
\cite{Lee:1964is,Kinoshita:1962ur}. In the soft photon limit the real
photon emission (bremsstrahlung) contribution factorizes to the born
cross section \cite{Bonneau:1971mk} and we include the corresponding
factor in our definition of $\sigma_{EM}$ for photon energies
$E(\gamma)\leq 0.1 \sqrt{s}$, again following \cite{Denner:1992bc}.
Hard photon corrections at the given order of the perturbative
expansion are insensitive to new physics contributions and we do not
consider them in the following; they can be vetoed in the actual
analysis.  To get a feeling for the magnitude of contributions coming
from loops of heavy particles we can also investigate the weak
corrections alone by defining
\be 
\sigma_W
= \sigma_0 + \sigma_{Weak} ~~, 
\ee 
with the electromagnetic corrections omitted.  We have verified 
the cancellation of UV divergences for both the weak and
electromagnetic corrections and also the cancellation of IR
divergences in the electromagnetic corrections.

Comparing our results with those presented in \cite{Denner:1992bc} we
find excellent agreement.  Updating with the known top-quark and Higgs
mass (which we set to $m_h = 125$~GeV throughout) we calculate the
weak corrections within the SM at a future $250$~GeV lepton collider.
In \Fig{fig:smtotcross} we plot the total associated production cross
section as a function of $\sqrt{s}$ (which peaks near $\sqrt{s}
\approx 250$~GeV) and in \Fig{fig:smdiff} we plot the differential
cross section at $\sqrt{s} = 250$~GeV.  The weak corrections are
typically small and positive for all polarizations and scattering
angles.  The electromagnetic corrections are slightly larger and
negative, thus, when combined, the weak and electromagnetic
corrections lead to an overall reduction in the associated production
cross section.  The weak corrections remain below $5 \%$ of the
tree-level cross section, and for unpolarized $Z$-bosons constitute an
$\mathcal{O} (3 \%)$ correction at $\sqrt{s} = 250$~GeV.  Although
these corrections are small it is expected that the $hZZ$ coupling can
be determined to $\mathcal{O} (1\%)$ accuracy at a $250$~GeV lepton collider
with $250~{\text{fb}}^{-1}$ integrated luminosity \cite{Klute:2013cx},
and so weak radiative corrections can in principle be measured at such
a machine.

It is clear that SM weak radiative corrections, although small, are
large enough to be relevant to lepton collider Higgs studies.  Their
magnitude, $\mathcal{O} (3 \%)$, would lead one to expect corrections
arising due to perturbative BSM scenarios to be similar.

\subsection{Beyond the SM}
\label{sec:BSM}

We calculate these corrections following the same methods as for the
SM weak corrections and include all contributing diagrams, including
those that contribute to vector boson self energies.  We define the
BSM correction to the associated production cross section
\be 
\delta \sigma_{Zh} = \frac{\sigma_{BSM} (e^+ e^- \rightarrow Z
  h)-\sigma_{SM} (e^+ e^- \rightarrow Z h)}{\sigma_{0} (e^+ e^-
  \rightarrow Z h)} ~~,
\label{eq:deltsig}
\ee
and in a similar manner define
\be
R_{h\gamma \gamma} = \frac{\Gamma_{BSM} (h \rightarrow \gamma
  \gamma)} {\Gamma_{SM} (h \rightarrow \gamma \gamma)} =
\frac{\text{BR}_{BSM}(h\to\gamma\gamma)}{\text{BR}_{SM}(h\to\gamma\gamma)} 
\left[\frac{\Gamma^{\text{tot}}_{BSM}}{\Gamma^{\text{tot}}_{SM}}\right]^{-1}~~.
\label{eq:Rgamgam}
\ee
Recent ATLAS fits indicate total width constraints
$\Gamma^{\text{tot}}_{BSM}/\Gamma^{\text{tot}}_{SM}\lesssim 2.12$
\cite{newatlasferm}. While this number results from an overall
scaling factor of LHC measurements and will therefore be limited
systematically, the total Higgs width can be accessed at a $250$ GeV
lepton collider via a measurement of the $e^+e^-\to \nu \bar \nu h$
channel with subsequent decay $h\to W^+W^-$ alone at the 10\% level
\cite{duerig}.  We will assume throughout that all new fields are too
heavy to introduce important new Higgs decay channels, and hence
$\Gamma^{\text{tot}}_{BSM} = \Gamma^{\text{tot}}_{SM}$, such that
modifications of the diphoton decay rate arise solely due to BSM
contributions to the decay amplitude.

In general BSM contributions to $\delta \sigma_{Zh}$ and $R_{h\gamma
  \gamma}$ are correlated, however the form of this correlation is
model dependent; the dependence arising through the spins and
electroweak charges of the additional BSM fields.  For this reason we
consider this correlation in the two sample BSM scenarios described in
\Sec{sec:sample}.

\subsubsection{Vector-Like Fourth Generation Leptons}
\begin{figure}[!t]
  \centering
  \subfloat[]{\label{fig:deltasigma1a}\includegraphics[height=2.9in]{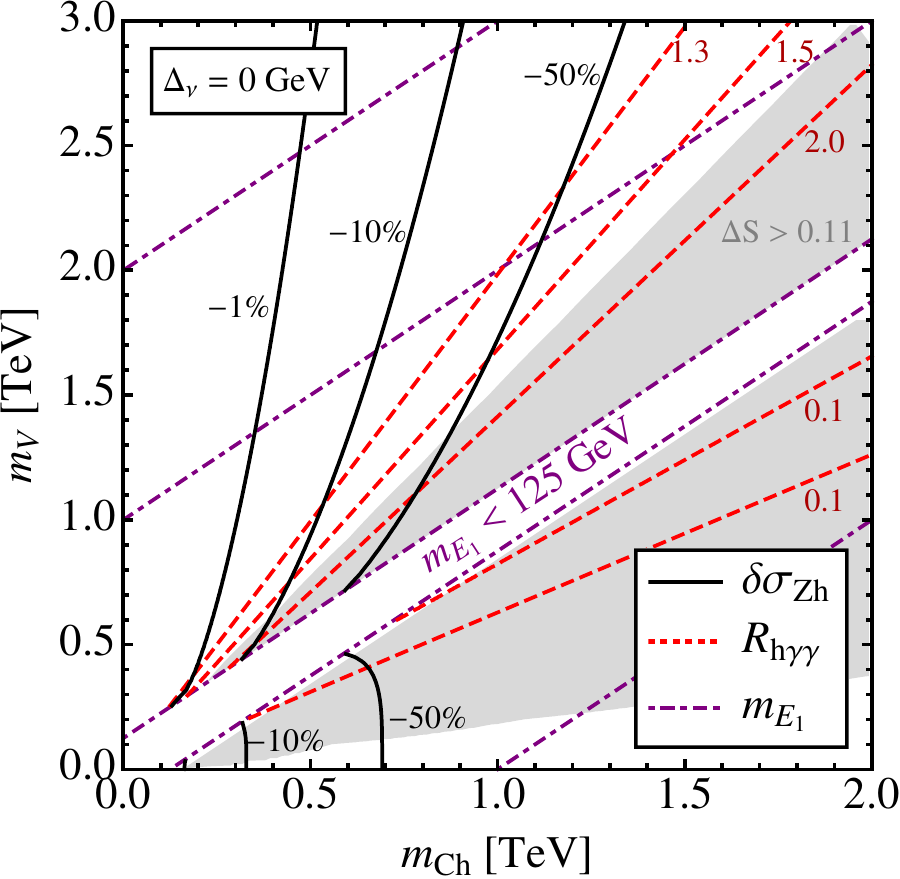}}   
  \hspace{0.1in}
  \subfloat[]{\label{fig:deltasigma1b}\includegraphics[height=2.9in]{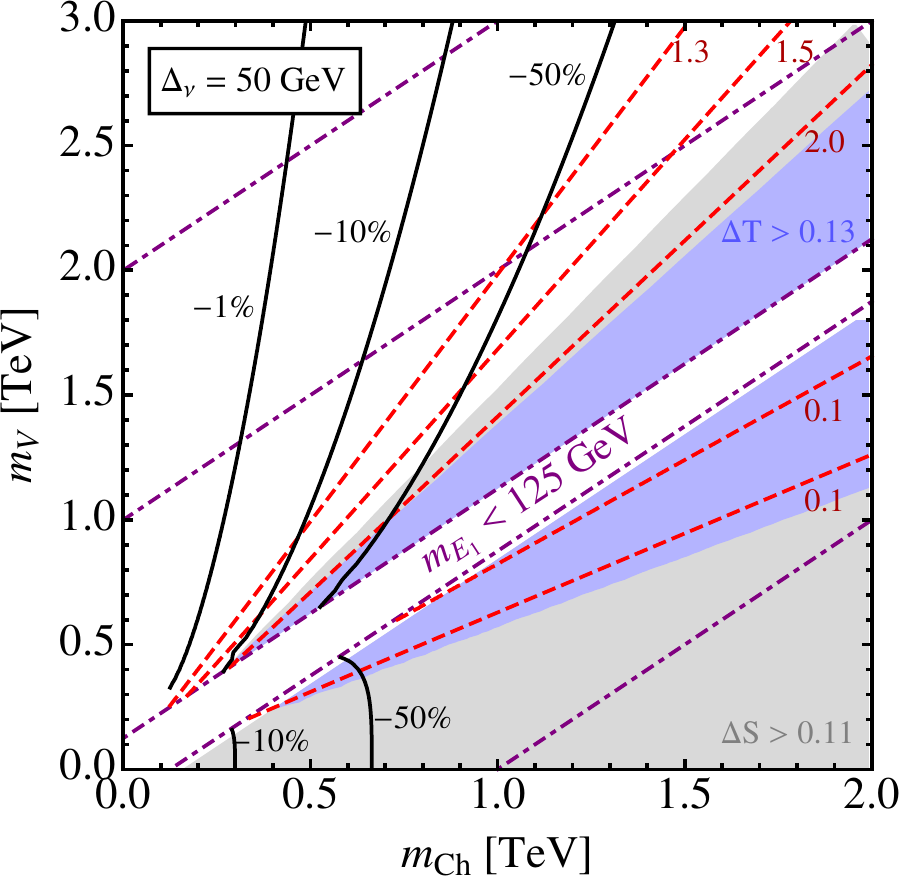}}
  \caption{Contours of the modified Higgs diphoton decay rate,
    $R_{h\gamma\gamma}$ (dashed red), and BSM corrections to the
    associated production cross section $\delta \sigma_{Zh}$ (solid
    black) as defined in \Sec{sec:BSM}, for the vector-like fourth
    lepton generation model.  Contours of the lightest charged fermion
    mass, $m_{E_1}$, are also shown (dotdashed purple) but not labelled.  For our
    restricted parameter choice we have $m_{E_1} = |m_V-m_{Ch}|$ so
    the lightest fermion mass can be read off from the intersection of
    these lines with the $x$ or $y$ axes.  (a) corresponds to the case
    where the Higgs contributions to charged and neutral fermion
    masses are equal, $\Delta_\nu = 0$~GeV, (see \Eq{eq:VL4hcoup}) and
    (b) to the case where the neutral fields couple more strongly to
    the Higgs than charged fermions, $\Delta_\nu = 50$~GeV.  Regions
    where precision electroweak corrections exceed measured values by
    $1\sigma$ are shaded, with $\Delta S$ in gray and $\Delta T$ in
    blue.  Parameter values where modifications to the diphoton rate
    are not significant ({\it e.g.}\ $\mathcal{O} (\lesssim 2 \sigma)$
    corresponding to the region to the left of the $R_{h\gamma\gamma}
    = 1.3$ contour) typically imply a reduction in the associated
    production which can comfortably exceed $1\%$ compared to the SM
    prediction.  This would be observable at a $250$~GeV lepton
    collider which can measure this cross section to $\mathcal{O}
    (1\%)$ accuracy.}
  \label{fig:VL4DeltaSigma}
\end{figure}

This simplified model, presented in \Sec{sec:VL4}, is defined by the
parameters $\{m_{Ch},m_V,\Delta_\nu \}$.  In each panel of
\Fig{fig:VL4DeltaSigma} we fix $\Delta_\nu$ to a specific value and
vary $m_{Ch}$ and $m_V$, plotting contours of the lightest charged
fermion mass, $m_{E_1}$, the Higgs diphoton decay rate
$R_{h\gamma\gamma}$, and corrections to the associated production
cross section $\delta \sigma_{Zh}$.  We also shade regions disfavored
by precision electroweak measurements.

Since modifications of the Higgs diphoton decays do not involve the
neutral fields, $R_{h\gamma\gamma}$ depends only on $m_{Ch}$ and
$m_V$.  It is clear that over the majority of allowed parameter space
corrections to $R_{h\gamma\gamma}$ are typically quite modest,
$\mathcal{O} (10\text{'s} \%) $ compared to the $\mathcal{O} (15 \%) $
model-independent experimental accuracy of the $h\gamma\gamma$
coupling.  Achieving an enhancement of the diphoton decay rate
requires $m_V>m_{Ch}$.  This might appear surprising since this
implies the dominant mass contribution does not come from the Higgs.
However, if the fermion mass comes solely from the Higgs the diphoton
decay rate never receives an enhancement since the contributions from
loops of fermions are always smaller in magnitude and opposite in sign
to the dominant amplitude from loops of $W$-bosons.\footnote{To
  achieve a large enhancement in this scenario one could always
  include multiple copies or larger charges, however such
  modifications are not within the definition of the model being
  studied here.}  If one increases $m_V$ then the lightest fermion can
be kept light by also increasing $m_{Ch}$.  This then allows for much
larger couplings to the Higgs for a given light fermion mass and hence
a larger contribution to the Higgs diphoton amplitude.

Considering \Fig{fig:VL4DeltaSigma} it is clear that within this model
BSM-NLO corrections typically lead to a reduction in the associated
production cross section of $1\%$ or greater, even in cases where
corrections to the diphoton rate are not significant.  With
sensitivity to the associated production cross section at
$\mathcal{O}(1\%)$ such deviations would be easily measurable at a
lepton collider, even if the Higgs diphoton rate is within $\mathcal{O}(1-2)
\sigma$ of the SM value.

Although not shown here we have also investigated the dependence of
the cross section corrections on $Z$-boson polarizations and
scattering angles.  The corrections to the associated production cross
section do not discriminate much between longitudinal and transverse
$Z$-boson polarizations, with similar corrections to both.
Furthermore, the corrections are also largely independent of the
scattering angle.

In conclusion, if the Higgs is coupled to additional heavy charged
fermions then, even if these fermions remain beyond direct collider
reach, and even if their corrections to the diphoton decay rate are
not significant, their presence should be evident at a lepton collider
through modifications to the associated production cross section.

\subsubsection{New Electroweak Scalars}

\begin{figure}[!t]
  \centering
  \includegraphics[height=2.6in]{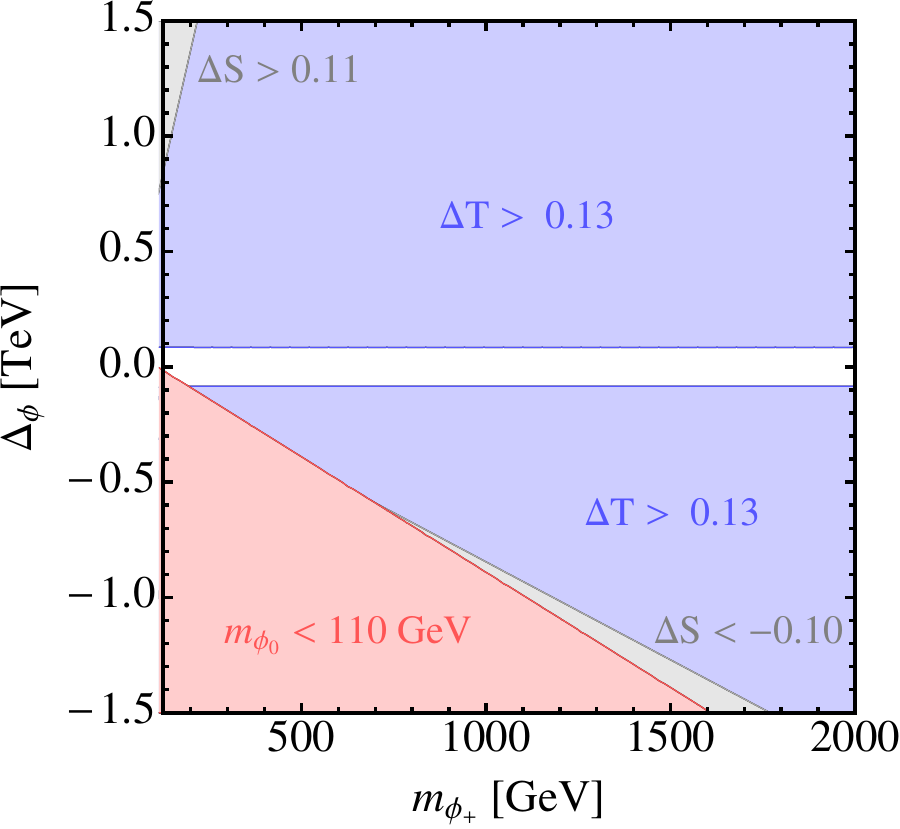} 
  \caption{Precision electroweak constraints on the electroweak scalar
    model.  By far the strongest constraint comes from the
    $T$-parameter and consistency at $1 \sigma$ with precision
    electroweak measurements requires $|\Delta_\phi| < 84$~GeV.
    Inconsistency with the $S$-parameter only arises in small wedges
    of parameter space.  We also require that the neutral scalar has
    mass greater than $m_{\phi_0}>110$~GeV to ensure consistency with
    LEP bounds.}
\label{fig:scalarSTU}
\end{figure}

This model is defined by the parameters
$\{m_{\phi_+},A_{\phi_+},\Delta_\phi \}$ and for a particular choice
of these parameters both $\delta \sigma_{Zh}$ and $R_{h\gamma \gamma}$
are completely determined.  As shown in \Fig{fig:scalarSTU}
the charged-neutral scalar mass splitting $\Delta_\phi$ is
constrained by precision electroweak measurements to be $|\Delta_\phi|
< 84$~GeV.  With this constraint imposed this model is consistent with
precision electroweak measurements for the full range of $m_{\phi_+}$.

\begin{figure}[!t]
  \centering
  \subfloat[]{\label{fig:deltasigmac1a}\includegraphics[height=2.8in]{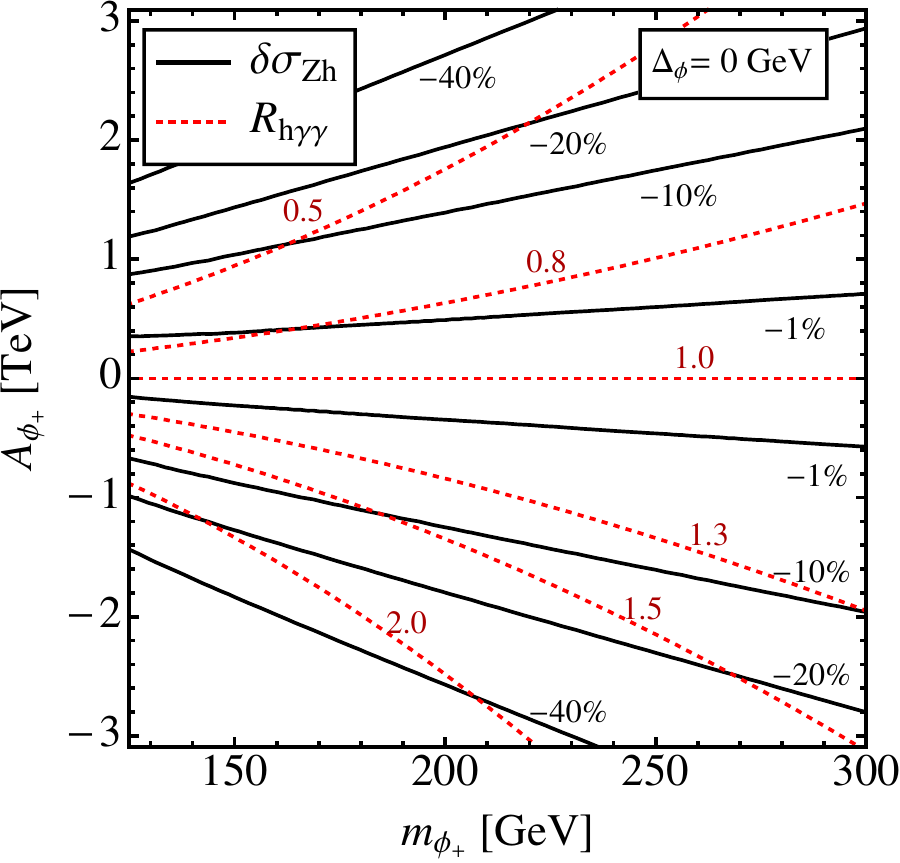}}
  \hspace{0.1in}
  \subfloat[]{\label{fig:deltasigmac1b}\includegraphics[height=2.8in]{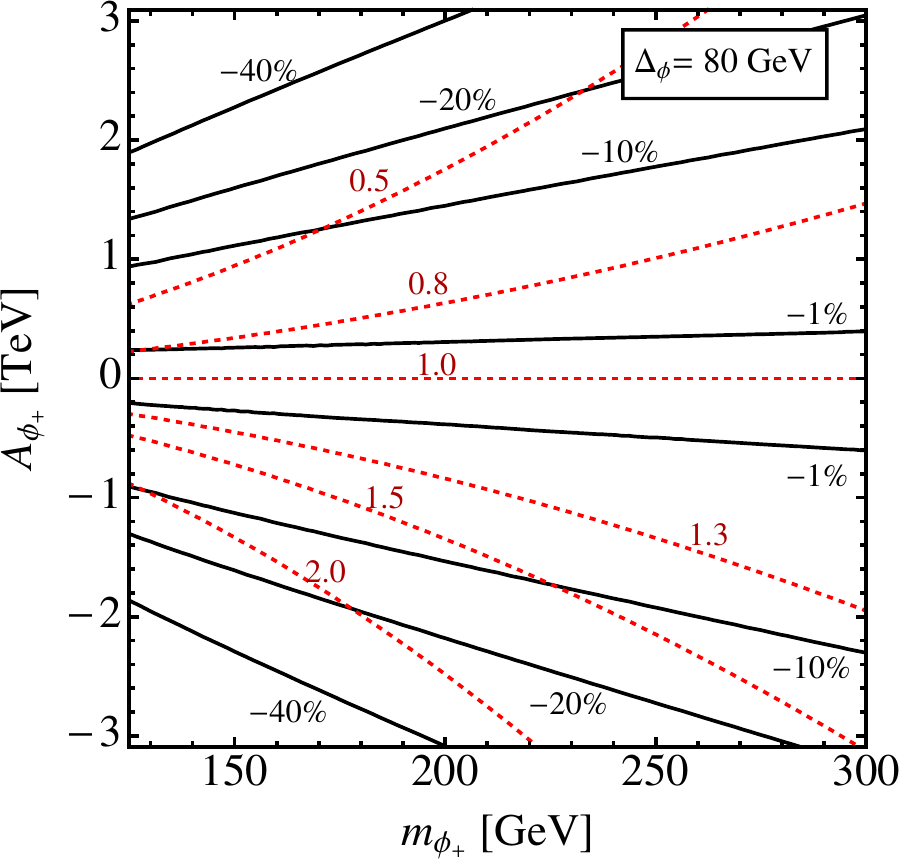}}
  \caption{Contours of $\delta \sigma_{Zh}$ (see \Eq{eq:deltsig}) in
    solid black, and $R_{h\gamma \gamma}$ (see \Eq{eq:Rgamgam}) in
    dotted red, for the charged scalar model with varying scalar
    masses and Higgs couplings.  Corrections to the associated
    production cross section are often $\mathcal{O} (\text{few}) \%$
    even when corrections to the diphoton rate are not significant
    $|R_{h\gamma \gamma} -1| \lesssim 0.3$, (corresponding to an
    $\mathcal{O} (2 \sigma)$ deviation in $R_{h\gamma \gamma}$).  With
    sensitivity to the associated production cross section at the
    $\mathcal{O} (1\%) $ level these modifications to associate
    production could be observed at a linear collider.}
  \label{fig:DeltaSigma}
\end{figure}

In \Fig{fig:DeltaSigma} we plot contours of $\delta \sigma_{Zh}$ and
$R_{h\gamma \gamma}$ for fixed values of $\Delta_\phi$ and varying
values of $\{m_{\phi_+},A_{\phi_+} \}$.  From these plots it is clear
that, within this particular model, it is possible to generate
deviations in the associated production cross section which are larger
than the $\mathcal{O} (1\%)$ achievable experimental resolution.
Again, although not shown here, the corrections to the associated
production cross section do not discriminate greatly between
longitudinal and transverse $Z$-boson polarizations and are also
largely independent of the scattering angle.

Hence, similarly to the model with vector-like fermions, if the Higgs
is coupled to new charged scalars close to the weak scale then, even
in the case where these scalars remain beyond direct reach of the LHC
and lepton colliders, there is still the potential for measuring
significant deviations in the associated production cross section at a
linear collider, assuming the cross section can be measured to an
accuracy of $1\%$.

\subsubsection{Dominant Corrections in the Scalar Model}
\label{sec:domi}
In \Fig{fig:DeltaSigma} one can see that the one-loop BSM corrections,
$\delta \sigma_{Zh}$, are, to a reasonable approximation, a quadratic
function of the trilinear coupling $A_{\phi_+}$.  Of all contributing
diagrams, only the scalar loops which contribute to the Higgs
wavefunction renormalization have a quadratic dependence on the scalar
coupling.  This wavefunction renormalization then feeds into the $hZZ$
vertex through a field redefinition to give canonically normalized
external states.  This contribution to the effective vertex is finite
and if we only include it then modifications to the associated
production cross section can be simply written as
\be 
\delta \sigma_{Zh} \approx \frac{1}{16 \pi^2 m_h^2} \left(
  A_{\phi_+}^2 G(\tau_+) + A_{\phi_0}^2 G(\tau_0) \right)~~,
\label{eq:approxloop}
\ee
where $\tau_+ = (m_h/2 m_{\phi_+})^2$ and $\tau_0 = (m_h/2
m_{\phi_0})^2$, which are commonly found quantities in loop integrals.
$A_{\phi_+}$ is the trilinear coupling between the Higgs and charged
scalars, as defined in \Sec{sec:Scalars}, and $A_{\phi_0}$ is the
trilinear coupling between the Higgs and neutral scalars, which can
also be written as
\be
A_{\phi_0} = A_{\phi_+}+\frac{2}{v_h} (m_{\phi_0}^2-m_{\phi_+}^2)  ~~,
\ee
where $v_h=246$ GeV is the usual Higgs vev.  Finally we define the
function
\be 
G(\tau) = 1+\frac{1}{4 \sqrt{\tau (\tau-1)}} \left( \log \left( 1-
    2 \tau - 2 \sqrt{\tau (\tau-1)} \right) - \log \left( 1- 2 \tau +2
    \sqrt{\tau (\tau-1)} \right) \right) ~~,
\label{eq:scalampf}
\ee 
which is related to the derivative of the scalar two-point function.
Since $0<\tau_{+,0} <1$ for the cases of interest the 
denominator and arguments of the logarithms in \Eq{eq:scalampf} are 
complex, however the final result is real.

Comparing with the full one-loop corrections we find that whenever the
scalar coupling is large, $A_{\phi_{+,0}} \gtrsim v_h$, then
\Eq{eq:approxloop} gives a reasonably good approximation to the full
expression.  It is interesting that the dominant one-loop correction
is entirely independent of the electroweak charges of the scalars,
demonstrating that the associated production cross section is a
sensitive collider-probe of scenarios where the Higgs is coupled to
neutral scalars via the Higgs-portal
\cite{Silveira:1985rk,McDonald:1993ex,Burgess:2000yq,Davoudiasl:2004be,Patt:2006fw,Schabinger:2005ei,Englert:2011aa,Binoth:1996au}.
This is especially interesting as it applies even if the scalars are
too heavy to introduce invisible decays.  Thus, if there are no
modifications to the Higgs diphoton decay rate, and if the Higgs has
no invisible decay width, the associated production cross section can
be used to probe the influence of additional neutral states at one
loop.\footnote{As this leads to an almost universal correction to all
  Higgs couplings evidence for such a scenario might also
  arise through studying other Higgs production channels.}

\section{BSM Electroweak Radiative Corrections at the LHC}
\label{sec:radLHC}
We now turn to a discussion of the extended electroweak corrections at
the LHC. Obviously, the electroweak NLO corrections
\cite{Ciccolini:2003jy} are not as important as the involved higher
order and re-summed QCD corrections
\cite{Ohnemus:1992bd,Baer:1992vx,Han:1991ia,Brein:2003wg,Hamberg:1990np,Dawson:2012gs,Ciccolini:2003jy}
which currently give rise to an ${\cal{O}}(10\%)$ residual
perturbative uncertainty \cite{Dittmaier:2011ti}. It is however not
unreasonable to expect that this number will improve over time when
more data becomes available and theoretical development is stimulated,
such that potentially large electroweak modifications can be resolved.

Technically, we proceed along the lines of Sec.~\ref{sec:lincol} to
calculate the electroweak NLO corrections. We use the CTEQ6l1 parton
distribution functions (pdfs) \cite{Pumplin:2002vw} following
Ref.~\cite{Ciccolini:2003jy} within the {\sc{Vbfnlo}} framework
\cite{Arnold:2008rz} to handle the pdf-folding and phase space
integration for $\Delta E(\gamma)\leq 0.1~\sqrt{\hat{s}}$ ($\hat{s}$
denotes the partonic $s$). We have again checked the cancellations of
both UV and IR singularities and we find excellent agreement comparing
to integrated results from {\sc{FormCalc}}/{\sc{LoopTools}} which
validates our phase space integration. Our leading order matrix
element has been cross checked against
{\sc{MadGraph}}~\cite{Alwall:2007st} for individual phase space
points.\footnote{Electroweak processes also typically depend on the
  electroweak scheme that is used to construct the parameters of the
  electroweak sector from a minimal set of
  observables~\cite{Ciccolini:2003jy}. This dependence can be large
  but is not important for us since it does not involve the BSM
  contributions.}

\begin{figure}
  \centering
  \subfloat[]{\label{fig:pt}\includegraphics[height=2.35in]{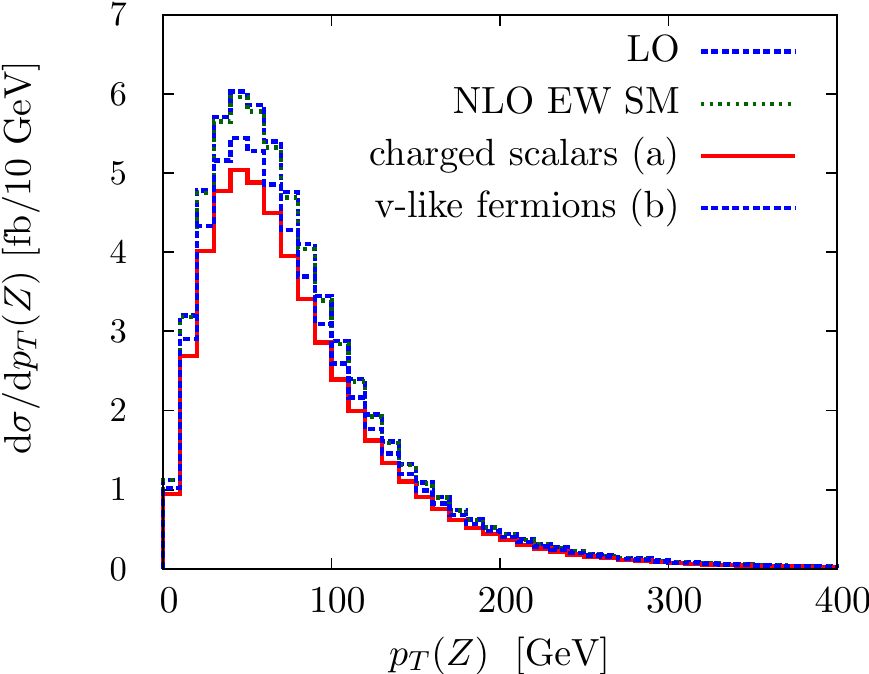}}   
  \hspace{0.2in}
  \subfloat[]{\label{fig:y}\includegraphics[height=2.35in]{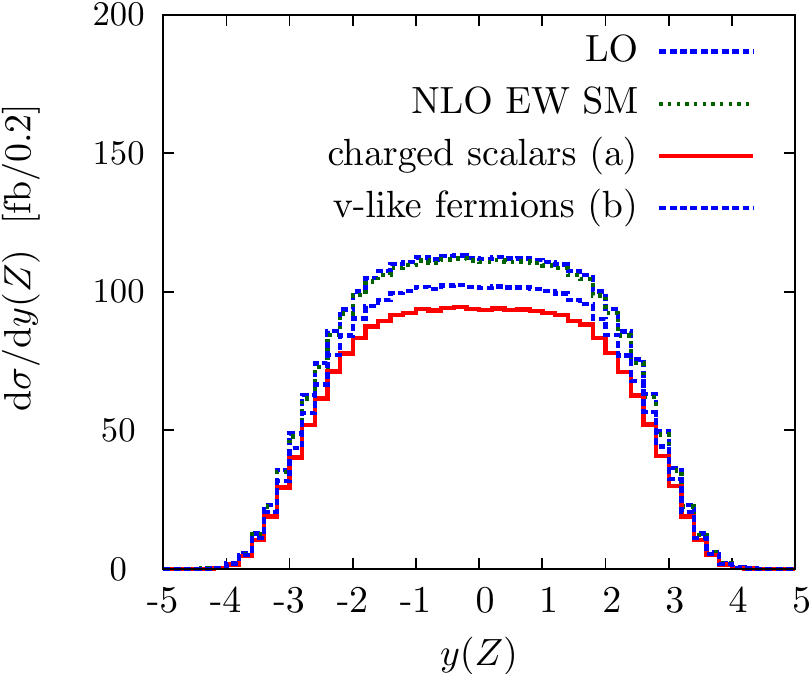}}
  \caption{(a) transverse momentum, and (b) rapidity, differential
    distributions of the $Z$ boson in $pp\to hZ$ production at the LHC
    $\sqrt{s}=14~\TeV$. We show leading order (LO) distributions as
    well as electroweak corrections in the SM and the BSM models
    discussed in the previous sections.  The BSM parameter choices are
    given in \Eq{eq:scena} and \Eq{eq:scenb}.  The integrated cross
    section receives corrections $\delta \sigma_{Zh} \simeq -15\%$ in
    the charged scalar model and $\delta \sigma_{Zh} \simeq -8\%$ in
    the vector-like fermion model.  The correction relative to the SM
    value, as a function of $p_T$, is shown separately in
    \Fig{fig:psd}.}
    \label{fig:LHCass}
\end{figure}

As done for the $e^+e^-$ case we regulate the soft photon divergencies
with a slicing parameter which is equivalent to dimensional
regularization \cite{Baur:1998kt}, and we do not consider either
initial state photon contributions or hard photon
corrections.\footnote{In principle, the presence of collinear mass
  singularities requires extending the list of subprocesses to
  photon-induced processes and the DGLAP kernels accordingly. Such pdf
  sets exist, see {\it e.g.}\ Ref.~\cite{Martin:2004dh}, but their
  contribution to the present process is known to be negligible $<1\%$
  for our purposes \cite{Kripfganz:1988bd,Spiesberger:1994dm}. Note
  that none of these contributions is sensitive to the BSM extension
  that we study. The full treatment will not change our results;
  especially the adoption of Eq.~\eqref{eq:deltsig} to the LHC
  situation is insensitive to all non-BSM effects.}
  
In Fig.~\ref{fig:LHCass} we show differential distributions for the SM
and our BSM scenarios. For a $2\to 2$ scattering process the most
relevant distributions are transverse momentum $p_T$ and rapidity $y$.
Dependence on the azimuthal angle, $\Phi$, is trivial. The remaining
collider observables follow as functions of $(p_T,y,\Phi)$ and the
involved masses $m_Z,m_h$.

The charged scalar mass scenarios we will study in the following are
given by
\begin{align}
  \text{(a)}\phantom{'} \qquad & (m_{\phi_+}~A_{\phi_+},~\Delta_\phi)=(250~\GeV ,
  -2.0~\TeV,~80~\GeV)~~,\label{eq:scena} \\
  \text{(a}'\text{)} \qquad &  (m_{\phi_+}~A_{\phi_+},~\Delta_\phi)=(250~\GeV ,
  +0.7~\TeV,~80~\GeV)~~,
\end{align}
and we choose a parameter point for the vector-like lepton scenario
\begin{equation}
  \text{(b)}  \qquad (m_{Ch},m_V,\Delta_\nu)=(500~\GeV,~1000~\GeV,~0~\GeV)~~.
  \label{eq:scenb}
\end{equation}
These parameter points are motivated from our results of
Sec.~\ref{sec:BSM},
\Figs{fig:VL4DeltaSigma}{fig:DeltaSigma}. We choose $\text{(a}\text{)}$ and
$\text{(b}\text{)}$ in the lepton collider
$\{R_{h\gamma\gamma},\sigma(hZ)\}$ correlation such that the diphoton
decay rate takes an extreme value $R_{h\gamma\gamma}\simeq 1.5$, while
$\text{(a}'\text{)}$ is a parameter point with a $\sim 20\%$ decrease
in 
\begin{wrapfigure}[19]{r}{7.5cm}
  \centering
  \includegraphics[height=2.35in]{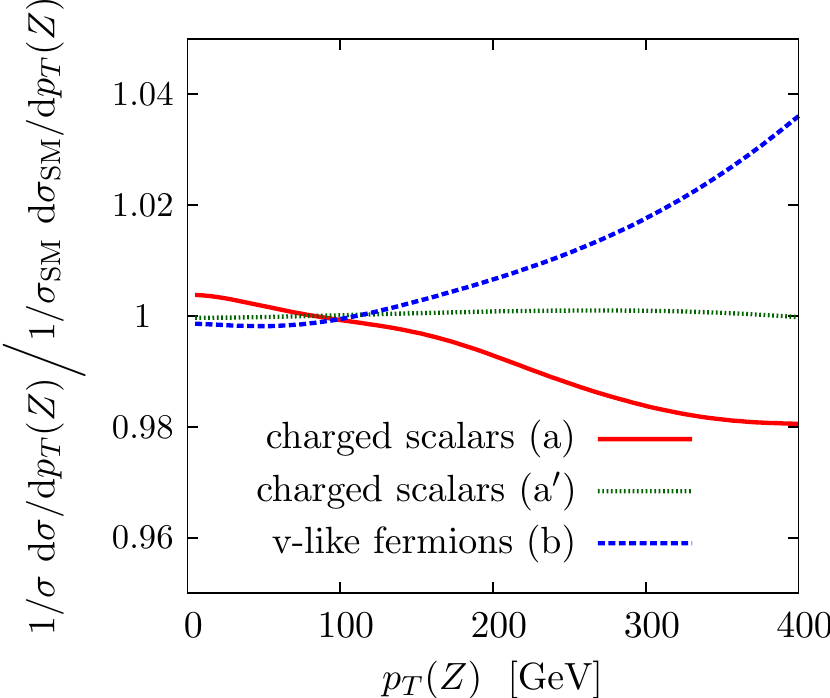}
  \parbox{7cm}{ \vspace{0.2cm} \caption{\label{fig:psd} Phase space
      dependence of the BSM electroweak corrections of the considered
      models and parameter points as a function of the $Z$ boson
      transverse momentum.}
  }
\end{wrapfigure}
$R_{h\gamma\gamma}$ compared to the SM.\footnote{The fourth
  generation vector-like lepton model does not admit
  $R_{h\gamma\gamma}<1$ due to electroweak precision constraints}

We find LHC BSM cross sections $\delta\sigma(hZ)\simeq -15\%,-8\%$ for
the parameter and model choices $\text{(a}\text{)}$ and
$\text{(b}\text{)}$, respectively. For $\text{(a}'\text{)}$ we obtain
$\delta\sigma(hZ)\simeq -1\% $. These numbers qualitatively reproduce
the lepton collider results of Sec.~\ref{sec:BSM} (we remind the
reader that $\delta\sigma$ is mostly driven by the BSM corrections).
As in the lepton collider case, the scalar extension shows a larger
deviation as compared to the fermion model. The scalar model is also
less constrained by electroweak precision results, which follows from
a formal decoupling of the non-oblique corrections (see also
Sec.~\ref{sec:domi}).

In Fig.~\ref{fig:psd} we show the phase space dependence in terms of
the $Z$ boson transverse momentum, which is sensitive to the BSM
contributions, especially when we start probing the new physics mass
scale. This justifies our choice to work in concrete models as
compared to effective theories mentioned in the introduction. This
effect is more clearly visible in Fig.~\ref{fig:psd}. The total cross
section is dominated from low $p_T$ configurations, yet for analyses
where associated production is accompanied with a high $p_T$ cut, such
as in $h\to b \overline{b}$ where $p_{T}(Z)\simeq 150~\GeV$, we
observe a turn-on of the new physics contribution which further
decreases the cross section compared to the SM expectation on the
inclusive level. The phase space dependence of the correction,
however, is highly dependent on the model. The overall effect is small
and in the percent range, but should nonetheless be considered in
precision studies that aim to determine Higgs properties at the LHC.
Within the context of the charged scalar model the parameter point ($\text{a}'$) does not result in important
effects on LHC associated production.

We have only discussed $pp\to hZ$ but expect our results should generalize
to $pp\to h W^\pm$ qualitatively as in the SM
\cite{Ciccolini:2003jy}, and so these results also have 
relevance for the $pp\to hW\to WWW$ channel \cite{CMS-PAS-HIG-12-039}.

\section{Conclusions}
\label{sec:Conclusions}
In this work we have investigated the higher order electroweak
corrections to associated Higgs production at a future lepton collider
and the LHC in electroweak extensions of the Higgs sector. Because
higher order corrections are model and phase space dependent, we have
studied them within two simple BSM scenarios which
involve vector-like leptons and charged scalars. Our results are of
importance for the consideration of Higgs physics at a future lepton
collider since couplings and branching ratios can be measured to high
precision. Reconciling the new physics contributions with current
limits from LEP measurements we find regions of parameter space with
${\cal{O}}(10\%)$ or larger deviations of the Higgs associated
production cross section, in comparison to the SM expectation.  These
corrections are much larger than the expected lepton collider
precision, and are possible even when corrections to the diphoton rate are within $\mathcal{O} (1-2)
\sigma$ of the expected precision.

We find similar corrections for LHC collisions. The phase space
dependence is typically such that the integrated BSM cross section
modifications are dominated by soft events, but high $p_T$ events
exhibit the largest deviations. These corrections, when compared to
the SM, can again be ${\cal{O}}(10\%)$, which might be resolvable for
high luminosity studies. The phase space dependence of the BSM
contribution then becomes a non-negligible parameter in precision
cross section and coupling extractions, which, depending on the data,
might challenge an effective field theory based analysis as {\it de
  facto} considered in Ref.~\cite{ATLAS-CONF-2012-127}.

While signals for any BSM physics at the LHC remain elusive it is
increasingly important to explore the full breadth of potential experimental BSM
signatures.  Although the direct observation of new BSM
states still remains a possibility, it may be that indirect evidence
in the form of anomalous properties of SM states and processes might
also uncover new physics.  Since many well-motivated BSM scenarios are
concerned with the Higgs sector, the newly discovered Higgs then
becomes, perhaps, the strongest candidate for indirect hints of new
physics and it is timely to understand the full complement of possible
BSM modifications to its properties.  This effort is already underway
with intense study in various respects, however this work highlights
that as we progress into the Higgs precision era it is important that
theoretical progress in SM precision Higgs calculations comes
accompanied by BSM precision Higgs calculations.
The scope of BSM precision Higgs physics is extensive and in this work
we have only considered corrections to Higgs associated production in
two simplified models.  However, this has demonstrated that if the
Higgs is coupled to new electroweak states then there exist regions of
parameter space where modifications to $h\to\gamma\gamma$ rates may
remain below statistical significance, while corrections to associated
production, particularly at a lepton collider, may in fact lead to the
most significant evidence for a modified Higgs sector.

\section*{Acknowledgements}
We thank Markus Klute for conversations and for comments on an early
version of the draft.  M.M. has benefitted from conversations with
Jesse Thaler and Itay Yavin.  C.E. acknowledges funding by the Durham
International Junior Research Fellowship scheme.  M.M. is supported by
a Simons Postdoctoral Fellowship and by the U.S. Department of Energy
(DOE) under cooperative research agreement DE-FG02-05ER-41360.

\end{document}